\documentclass[graybox]{svmult}

\usepackage{mathptmx}           
\usepackage{helvet}                 
\usepackage{courier}                

\usepackage{makeidx}              
\usepackage{graphicx}              
\usepackage{multicol}               

\usepackage{url}
\usepackage{amsmath,amssymb}
\usepackage{braket}

\usepackage{subcaption}
\newcommand{\mtx}[1]{\boldsymbol{\sf #1}}

\makeindex

\begin{document}

\title*{Quantum decoherence emulated in a classical device}
\author{Brian R. La Cour, Corey I. Ostrove, Michael J. Starkey, Granville E. Ott}
\institute{
Brian R. La Cour, Corey I. Ostrove, Michael J. Starkey, Granville E. Ott \at Applied Research Laboratories, The University of Texas at Austin \\ P. O. Box. 8029, Austin, Texas 78713-8029, USA \\
\email{blacour@arlut.utexas.edu}
}

\maketitle

\abstract{We demonstrate that a classical emulation of quantum gate operations, here represented by an actual analog electronic device, can be modeled accurately as a quantum operation in terms of a universal set of Pauli operators.  This observation raises the possibility that quantum error correction methods may be applied to classical systems to improve fault tolerance.}

\section{Introduction}
\label{sec:intro}

Decoherence in quantum systems arises as an inevitable consequence of interactions with the environment.  Theoretically, it has been used to understand the measurement problem and provide a gateway to the classical world \cite{joos2013decoherence}.  From a practical perspective, decoherence poses a challenge to developing large-scale quantum computers, whose efficacy degrades with the loss of coherence.  Quantum error correction meets this challenge by providing a scalable means of correcting a continuum of possible errors, representable by a finite set of operators, using only a discrete number of components \cite{shor1995scheme,gottesman2009introduction}.  In this paper, we will investigate whether classical analog systems can exhibit similar behavior.

We shall begin in Sec.\ \ref{sec:qed} with a description of a classical emulation of a gate-based quantum computer.  There, we describe how arbitray quantum states may be represented by classical signals conforming to the mathematical tensor-product structure of a multi-qubit Hilbert space.  We go on to describe how one- and two-qubit gate operations can be performed on these representative states and even how quantum measurements can be faithfully emulated.  Finally, describe briefly a hardware implementation of a device capable of emulating a two-qubit quantum computer.

We go on, in Sec.\ \ref{sec:qst}, to describe how one may use such a device to perform quantum state tomography and, therefore, infer the equivalent mixed quantum state from an ensemble of imperfect state preparations.  In Sec.\ \ref{sec:qpt} we extend our investigation to modeling gate operations in our classical device in terms of quantum operations.  This provides the natural mathematical framework for studying classical performance degradation in terms of quantum decoherence.  By applying a sequence of gate operations iteratively, we are able to measure the systematic falloff in performance of the device and model it as a parameterized quantum channel.  Our conclusions are summarized in Sec.\ \ref{sec:end}.


\section{Classical Emulation of a Quantum Device}
\label{sec:qed}

In previous work, we have described the use of classical signals and analog electronics to emulate or ``mimic'' the behavior of a gate-based quantum computer \cite{LaCour&Ott2015}.  The basic idea is quite simple.  Given an abstract Hilbert space $\mathcal{H}$ used to represent an $n$-qubit quantum state $\ket{\psi} \in \mathcal{H}$, we seek a classical representation of these abstract mathematical objects.  Many possibilities suggest themselves.  The one we shall choose is motivated by our desire to easily perform operations on it.  To that end, we adopt sinusoidal analog signals as a convenient physical representation of a pure quantum state.

If we denote by $\ket{x}$, where $x \in \{0, \ldots, 2^n-1\}$, a basis function in the so-called computational basis of $\mathcal{H}$, then its classical representation is defined as follows.  Let $[x_{n-1} \cdots x_0]$ be the little endian binary representation of $x$.  The corresponding classical representation is then written as a complex signal $\phi_x$ defined such that
\begin{equation}
\phi_x(t)  = \exp[(-1)^{x_{n-1}} i \omega_{n-1} t] \cdots \exp[(-1)^{x_{0}} i \omega_{0} t] \; ,
\end{equation}
where the frequencies $\omega_{n-1} > \cdots > \omega_0$ correspond to each of the $n$ qubits.  In particular, taking $\omega_k = 2^k \omega_0$ allows for a uniform spacing among the $2^n$ different combinations of sums and differences.  Clearly, the required bandwidth for such a representation grows exponentially with the number of qubits.  Linear combinations of basis signals provide a representation for an state in $\mathcal{H}$.  Thus, if $\braket{x|\phi} = \alpha_x$, then
\begin{equation}
\psi(t) = \sum_{x=0}^{2^n-1} \alpha_x \, \phi_x(t)
\end{equation}
provides a classical representation of the quantum state $\ket{\psi}$.  If $T = 2\pi/\omega_0$ is the period of the signal, then an inner product may be defined as
\begin{equation}
\braket{\phi_x|\psi} = \frac{1}{T} \int_0^T \phi_x(t)^* \psi(t) \, dt \; .
\end{equation}
Note that the set of computational basis signals forms an orthonormal basis.

Gate operations are performed using projection operations.  Given a quantum state $\ket{\psi}$ and a qubit $i$ to be addressed, we can formally decompose it into projections onto the subspaces in which the qubit is either 0 or 1.  Thus, we may write 
\begin{equation}
\ket{\psi} = \Pi_0^{(i)} \ket{\psi} + \Pi_1^{(i)} \ket{\psi} = \ket{0}_i \ket{\psi_0^{(i)}} + \ket{1}_i \ket{\psi_1^{(i)}} \; ,
\end{equation}
where $\ket{\psi_0^{(i)}}$ and $\ket{\psi_1^{(i)}}$ will be called partial projection states.  As described in Ref.\ \cite{LaCour&Ott2015}, the corresponding signals $\psi_0^{(i)}$ and $\psi_1^{(i)}$ may be produced from the signal $\psi$ using classical analog signal processing devices.  In this manner, the nonseparable subspace projections may be separated and operated upon individually.  Doing so allows one to perform a single-qubit gate operation $U$ on qubit $i$ by noting that
\begin{equation}
U_i \ket{\psi} = U\ket{0}_i \ket{\psi_0^{(i)}} + U\ket{1}_i \ket{\psi_1^{(i)}} \; .
\end{equation}
So, simple multiplication and addition of analog signals is all that is needed to effect this transformation.

Measurements are performed in a similar manner.  Given the partial projection signals $\psi_0^{(i)}$ and $\psi_1^{(i)}$ we may perform a measurement on qubit $i$ by first measuring the root-mean-square (RMS) voltages $v_0 = \|\psi_0^{(i)}\|$ and $v_1 = \|\psi_0^{(i)}\|$ of each signal and, from these, computing the probability
\begin{equation}
p = \frac{v_0^2}{v_0^2 + v_1^2} \; .
\end{equation}
A random variable $u_i \in [0,1]$, which serves the role of a hidden variable, is drawn such that $u_i \le p$ indicates an outcome of 0 and $u_i > p$ indicates an outcome of 1.  Upon measurement, the state ``collapses'' to form the new signal $\psi' \propto \psi_0^{(i)}$ or $\psi' \propto \psi_1^{(i)}$, depending upon the measurement outcome.  Additional qubits may be measured sequentially in this manner.  This procedure, then, faithfully reproduces the quantum statistics dictated by the Born rule.

We have implemented a small-scale quantum emulation device in hardware using breadboards and analog electronic components interfaced to a digital desktop computer.  The device can be operated in one- or two-qubit mode at frequencies of 1000 Hz and 2000 Hz, respectively.  Arbitary one-qubit gate operations can be performed as well as controlled gate operations using qubit 0 as the control and qubit 1 as the target.  Typical gate fidelities are found to be over 99\% \cite{LaCour&al2016}.  Measurement gates are performed using true-RMS voltage chips and digital switching.  Sequential gate operations can be performed using a software interface.  For example, a simple implementation of Deutch's algorithm can be programmed for an unknown Boolean function.  In practice, we find that the device is able to correctly identify whether the function is constant or balanced about 96\% of the time.  Since the algorithm should, ideally, produce the correct answer every time, the nonzero error rate must be due to device imperfections.  In the following sections, we will investigate whether these imperfections can be modeled as quantum decoherence.


\section{Application of Quantum State Tomography}
\label{sec:qst}

Using the representation and measurement procedure described in Sec.\ \ref{sec:qed}, we are able to use our device to prepare and measure any quantum state and observable of up to two qubits.  Of course, imperfections in the device itself can give only a limited approximation of the ideal mathematical operations.  This situation is quite similar to that found in actual quantum devices or experiments, and we may use similar tools to study it.

We have heretofore discussed the representation of \emph{pure} quantum states using our device, but the more general quantum description is that of a \emph{mixed} state.  Mixed quantum states may be thought of as an ensemble of pure states.  Equivalently, in our classical representation, they may be thought of as noisy signals.  For example, it can be shown that additive Gaussian white noise is equivalent to a mixture of the ideal state and a fully mixed state \cite{LaCour&Ostrove2017}.  Noise can also be added intentionally to reproduce certain quantum measurement effects, as was done with so-called dressed states \cite{LaCour&al2016}.  For the purposes of the present study, however, we are simply interested in estimating the equivalent quantum mixed state given a set of measurement outcomes.  This may be done using the technique of quantum state tomography (QST).

A general (mixed) $n$-qubit state $\rho$ may be decomposed into a basis of $4^n$ separable orthonormal operators using the Hilbert-Schmidt inner product, as follows:
\begin{equation}
\rho = \sum_{j_{n-1}=1}^{4} \cdots \sum_{j_0=1}^{4} \textcolor{black}{\mathrm{Tr}\left[ \rho \; \frac{\sigma^{(n-1)}_{j_{n-1}} \otimes \cdots \otimes \sigma^{(0)}_{j_0}}{2^n}  \right]} \frac{\sigma^{(n-1)}_{j_{n-1}} \otimes \cdots \otimes \sigma^{(0)}_{j_0}}{2^n} \; ,
\end{equation}
where
\begin{equation}
\sigma_1^{(k)} = \mtx{I}_k \; , \quad \sigma_2^{(k)} = \mtx{X}_k \; , \quad \sigma_3^{(k)} = \mtx{Y}_k \; , \quad \sigma_4^{(k)} = \mtx{Z}_k
\end{equation}
are the four Pauli spin operators applied to qubit $k$.  Since the trace represents an expectation value under the Born rule \cite{vonNeumann}, we may use this decomposition to empirically determine the quantum state by measuring each of the basis operators.

Let $\bar{B}_{j_{n-1},\ldots,j_0} \in \mathbb{R}$ denote the mean value obtained from a finite sample of measurements of the operator $\sigma^{(n-1)}_{j_{n-1}} \otimes \cdots \otimes \sigma^{(0)}_{j_0} / 2^n$.  From these results, one may estimate the quantum state to be
\begin{equation}
\bar{\rho} = \sum_{j_{n-1}=1}^{4} \cdots \sum_{j_0=1}^{4} \textcolor{black}{\bar{B}_{j_{n-1},\ldots,j_0}} \frac{\sigma^{(n-1)}_{j_{n-1}} \otimes \cdots \otimes \sigma^{(0)}_{j_0}}{2^n} \; .
\end{equation}
In practice, $\bar{\rho}$ will not be a valid quantum state, since the numerical coefficients are not guaranteed to yield an operator that is both positive definite and of unit trace.  A better procedure is to restrict one's search to valid quantum states and find the maximum likelihood estimate (MLE) of the quantum state, here denoted $\hat{\rho}$, that both satisfies this constraint and best fits the measured mean values.  To this end, we use an MLE  procedure developed by Altepeter, Jeffrey, and Kwiat under the assumption of independent Gaussian errors \cite{Kwiat2006}.

Once a valid estimate $\hat{\rho}$ of the quantum state is obtained, it may be compared to the ideal quantum state $\ket{\psi}$ by computing the fidelity $F$ of the former to the latter using the expression \cite{Jozsa1994}
\begin{equation}
F = \sqrt{\langle \psi | \hat{\rho} | \psi \rangle} \; .
\end{equation}
Note that $F$ is bounded between zero and one.  If $\hat{\rho} = \mtx{I} \otimes \cdots \otimes \mtx{I} / 2^n$ (a completely mixed state), then $F = 1/2^n$.  Thus, in practice we expect to find intermediate values of $F$ such that $1/2^n < F < 1$.

\begin{figure}
	\centering
	\begin{subfigure}[t]{.45\textwidth}
     \centerline{$\rho$}
		\includegraphics[width= .9\textwidth]{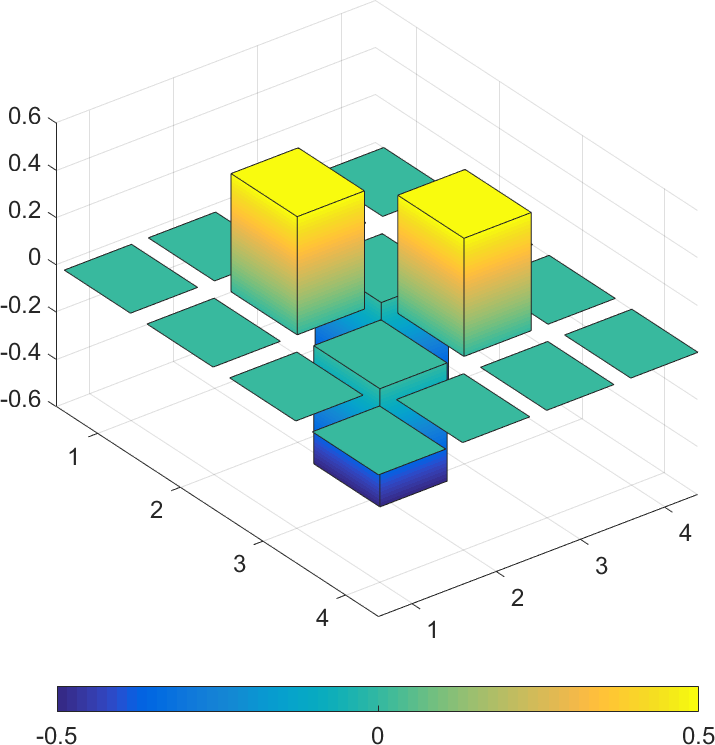}
		\caption{}
	\end{subfigure}
	\begin{subfigure}[t]{.45\textwidth}
     \centerline{$\hat{\rho}$}
		\includegraphics[width= .9 \textwidth]{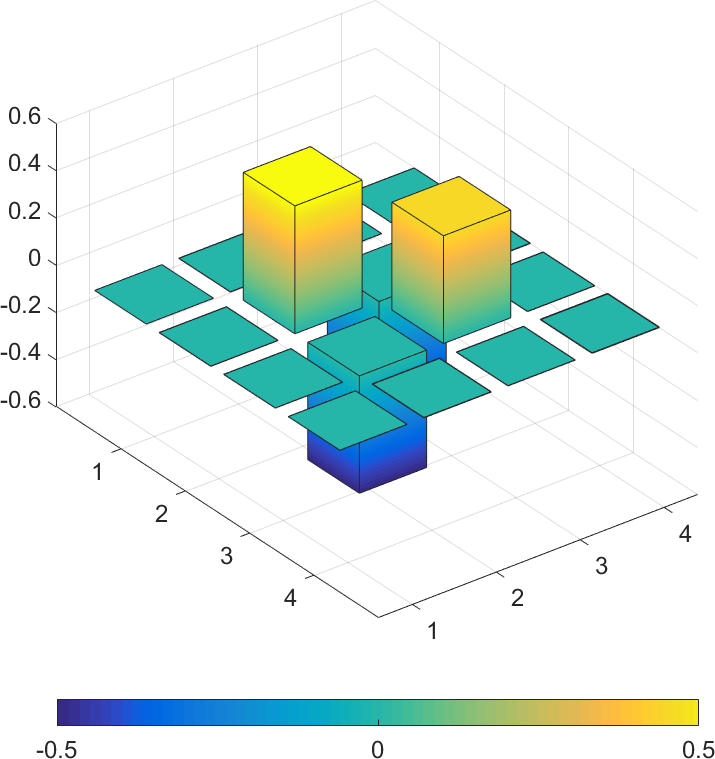}
		\caption{}
	\end{subfigure}
\caption{Cityscape plot of the ideal quantum state (left) and that inferred from quantum state tomography (right).  Only the real part of the matrix elements is shown.}
\label{fig:QST}
\end{figure}

As an example, we considered the pure entangled state $\ket{\psi} = \frac{1}{\sqrt{2}} [ \ket{01} - \ket{10} ]$.  So, in matrix form in the computational basis, the ideal quantum state is
\begin{equation}
\rho = \ket{\psi}\bra{\psi} = \frac{1}{2} \begin{pmatrix}
0 & 0 & 0 & 0 \\
0 & +1 & -1 & 0 \\
0 & -1 & +1 & 0 \\
0 & 0 & 0 & 0
\end{pmatrix} \; .
\end{equation}
The estimated quantum state was found to be
\begin{equation*}
\resizebox{\textwidth}{!}{$\hat{\rho}= \begin{pmatrix}
	0.0001 & -0.0011-0.0082i & \;0.0008+0.0075i & -0.0006+0.0004i   \\
	-0.0011+ 0.0082i & 0.5412 & -0.4968-0.0082i & 0.0019-0.0364i \\
	0.0008-0.0075i & -0.4968+0.0082i & 0.4562 & -0.0012+0.0335i \\
	-0.0006+0.0004i & 0.0019 + 0.0364i & -0.0012-0.0335i & 0.0025
	\end{pmatrix} \; ,$}
\end{equation*}
giving a fidelity of $F = 0.9978$.  The two states are shown graphically in Fig.\ \ref{fig:QST}.

\section{Application of Quantum Process Tomography}
\label{sec:qpt}

In quantum mechanics, the evolution of a closed system is given by a unitary transformation generated by the system's Hamiltonian, in accordance with the Schr\"odinger equation.  In practice, no system is ever truly isolated, and this can lead to apparent non-unitary evolution.  The formalism of quantum operations gives us a framework with which to characterize the behavior of open quantum systems and, in particular, decoherence \cite{Sudarshan1961,Mike&Ike}.  A quantum operation may be viewed as a superoperator on quantum states such that, if $\rho$ is the initial quantum state, then $\rho' = \mathcal{E}(\rho)$ is the state that results from some, possibly non-ideal, transformation.

We will make use of an equivalent formulation of quantum operations known as the operator-sum representation \cite{Stinespring1955}.  Using this formalism, the quantum operation $\mathcal{E}$ may be characterized by a discrete set of operators such that
\begin{equation}
\mathcal{E}(\rho) = \sum_k E_k \, \rho \, E_k^\dagger \; .
\end{equation}
The matrices $\{E_k\}$  are known as Kraus operators \cite{Kraus}.  A further decomposition of the Kraus operators may be performed in terms of, say, the Pauli operators, as was done for QST.

For our present purposes we will consider, for simplicity, only single-qubit states.  In this case, each Kraus operator may be written as a linear combination of the four Pauli operators, so that
\begin{equation}
E_k = \sum_{i=1}^{4} e_{k,i} \, \frac{\sigma_i}{2} \; .
\end{equation}
Using this representation, the quantum operation may be written as
\begin{equation}
\mathcal{E}(\rho) = \sum_k \left(\sum_{i=1}^{4} e_{k,i}\frac{\sigma_i}{2}\right) \rho \left(\sum_{j=1}^{4} e_{k,j}\frac{\sigma_j}{2}\right)^\dagger = \sum_{i=1}^{4} \sum_{j=1}^{4} \chi_{i,j} \; \sigma_i \, \rho \, \sigma_j^\dagger \; ,
\end{equation}
where
\begin{equation}
\chi_{i,j} = \sum_k e_{k,i} \, e_{k,j}^*
\end{equation}
are the elements of the $4\times4$ chi process matrix $\mtx{\chi}$ \cite{Choi1975,Chuang1997}.  Determination of the equivalent quantum operation therefore reduces to the problem of estimating the corresponding chi matrix.  This, in turn, may be accomplished using the techniques of quantum process tomography (QPT) \cite{Chuang1997,Poyatos1997,Bhandari2016}.

As an example, we performed QPT on a single-qubit identity gate using the procedure outlined in Ref.\ \cite{Mike&Ike} but modified to use a maximum likelihood QPT technique \cite{AGWhite2004mlqpt,Anis2012,Yuen-Zhou2014}.  We started by generating an ensemble of input states of the following form:
\begin{equation}
\ket{\psi_1} = \ket{0} \; , \;\; \ket{\psi_2} = \ket{1} \; , \;\; \ket{\psi_3} = \tfrac{1}{\sqrt{2}} [ \ket{0} + \ket{1} ] \; , \;\; \ket{\psi_4} = \tfrac{1}{\sqrt{2}} [ \ket{0} + i \ket{1} ] \; .
\end{equation}
The process, in this case an identity gate, was applied to the ensemble of states, which were then measured using this same basis.  The chi matrix was parameterized using a Cholesky factorization such that $\mtx{\chi} = \Delta \Delta^\dagger$, where $\Delta$ is a lower-triangular matrix with real, positive diagonal elements.  This form guarantees that the constraints of Hermiticity, trace preservation, and complete positivity are satisfied.  We then optimize over $\Delta$ with respect to the following likelihood function:
\begin{equation}
L(\Delta) = \frac{1}{2} \sum_{\alpha} \sum_{\beta} \frac{\left[ N_{\alpha,\beta} - C \sum_{i} \sum_{j} \bra{\psi_{\beta}} \sigma_i \ket{\phi_{\alpha}} \bra{\phi_{\alpha}} \sigma_j \ket{\psi_{\beta}} (\Delta \Delta^\dagger)_{i,j} \right]^2}{ C\sum_{i} \sum_{j} \bra{\psi_{\beta}} \sigma_i \ket{\phi_{\alpha}} \bra{\phi_{\alpha}} \sigma_j \ket{\psi_{\beta}} (\Delta \Delta^\dagger)_{i,j}} \; ,
\end{equation}
where $\ket{\phi_\alpha}$ and $\ket{\psi_\beta}$ are the input states and measurement settings, respectively, and $N_{\alpha , \beta}$ are the experimentally measured counts for the corresponding pair of input state and measurement settings.  The factor $C$ is the total number of such counts.

\begin{figure}
\centerline{\hspace{8em}$\mathrm{Re} \hat\chi$ \hfill $\mathrm{Im} \hat\chi$ \hspace{8em}}
\begin{center}
\includegraphics[width=\textwidth]{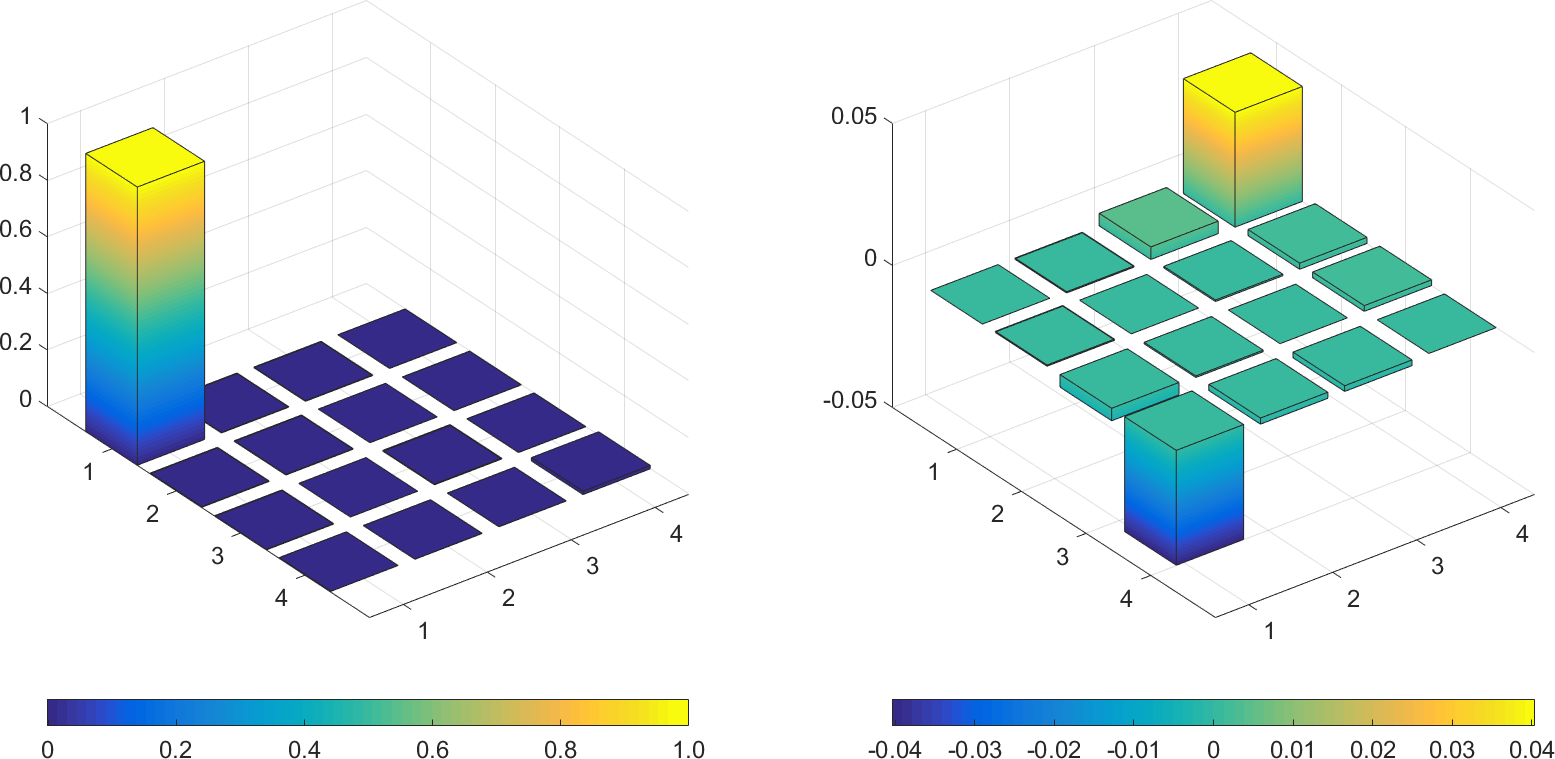}
\end{center}
\caption{Cityscape plot of the QPT results for a single application of the identity gate.  The real part of $\hat{\mtx{\chi}}$ is shown on the left, while the imaginary part of $\hat{\mtx{\chi}}$ is shown on the right.  Note that the two figures are shown on very different scales to illustrate the nonzero contributions to the estimate.}
\label{fig:QPT}
\end{figure}

The resulting estimated chi matrix, $\hat{\mtx{\chi}}$, is illustrated graphically in Fig.\ \ref{fig:QPT}.  Ideally, the matrix should be such that $\chi_{i,j} = \delta_{i,j}$, indicating that the quantum operation takes the simple form $\mathcal{E}(\rho) = \sigma_1 \, \rho \, \sigma_1 = \rho$.  Empirically, we do indeed find that $\hat{\chi}_{1,1}$ is nearly 1 and is, therefore, the dominant component.  A closer inspection reveals that there are other nonzero components.  In particular, the $\hat{\chi}_{1,j}$ and $\hat{\chi}_{j,1}$ components appear to have non-negligible imaginary terms.

The gate fidelity relative to an ideal unitary operator $U$ may be determined from the estimated quantum operation $\mathcal{E}$ according to the formula \cite{gilchrist2005distance}
\begin{equation}
F = \min_{\rho} \; \mathrm{Tr} \sqrt{\sqrt{\mathcal{E}(\rho)} \; U \rho U^\dagger \sqrt{\mathcal{E}(\rho)}} \; .
\label{eqn:gate_fidelity}
\end{equation}
%
For the present case, $U$ is the identity and $\mathcal{E}$ is estimated by the chi matrix $\hat{\mtx{\chi}}$, for which we find that $F = 0.9933$ for a single identity gate operation.  This is comparable to what was found earlier for the quantum state fidelity using QST and, so, these results appear to be consistent.

Ideally, the chi matrix should give a full characterization of the quantum process (in this case, a single application of the identity gate operation).  In particular, it should provide a means of forecasting the gate fidelity over multiple multiple iterations.  Indeed, if the initial quantum state is determined to be $\rho_0$, then the state after $n$ iterations, denoted $\rho_n$, is given iteratively by
\begin{equation}
\rho_n = \mathcal{E}(\rho_{n-1}) = \mathcal{E}( \cdots \mathcal{E}(\rho_0) \cdots ) \; .
\end{equation}
Consider the depolarizing channel with parameter $p \in [0,1]$, for which
\begin{equation}
\mathcal{E}(\rho) = (1-p) U \rho U^\dagger + p \mtx{I} \; .
\end{equation}
The fidelity of this channel is then
\begin{equation}
F = \sqrt{ 1 - \frac{2p}{3} } \; .
\end{equation}
The depolarizing channel is closed under multiple iterations, with an effective parameter $p_n$ after $n$ iterations of
\begin{equation}
p_n = \frac{3}{4} \left[ 1 - \left( 1 - \frac{4p}{3} \right)^n \right] \; ,
\end{equation}
yielding a cumulative fidelity of
\begin{equation}
F_n = \sqrt{1 - \frac{1}{2} \left[ 1 - \left( 1 - \frac{4p}{3} \right)^n \right] } \; .
\end{equation}

In addition to the gate fidelity defined in Eqn.\ (\ref{eqn:gate_fidelity}) we make use of an alternative benchmark known as the quantum process fidelity defined as \cite{gilchrist2005distance}
\begin{equation}
F_{\textrm{proc}}= \textrm{Tr}(\hat{\chi} uu^\dagger)
\label{eqn:process_fidelity}
\end{equation}
where $\hat{\chi}$ is the measured chi matrix and $uu^\dagger$ is the rank-one chi matrix for the ideal unitary transformation $U$.  The process fidelity has the benefit of being less computationally intensive to calculate as we need not perform the optimization step over input states. For the depolarizing channel, it may be interpreted as the probability that the ideal operation was performed.

\begin{figure}
\centering
\includegraphics[width=\textwidth]{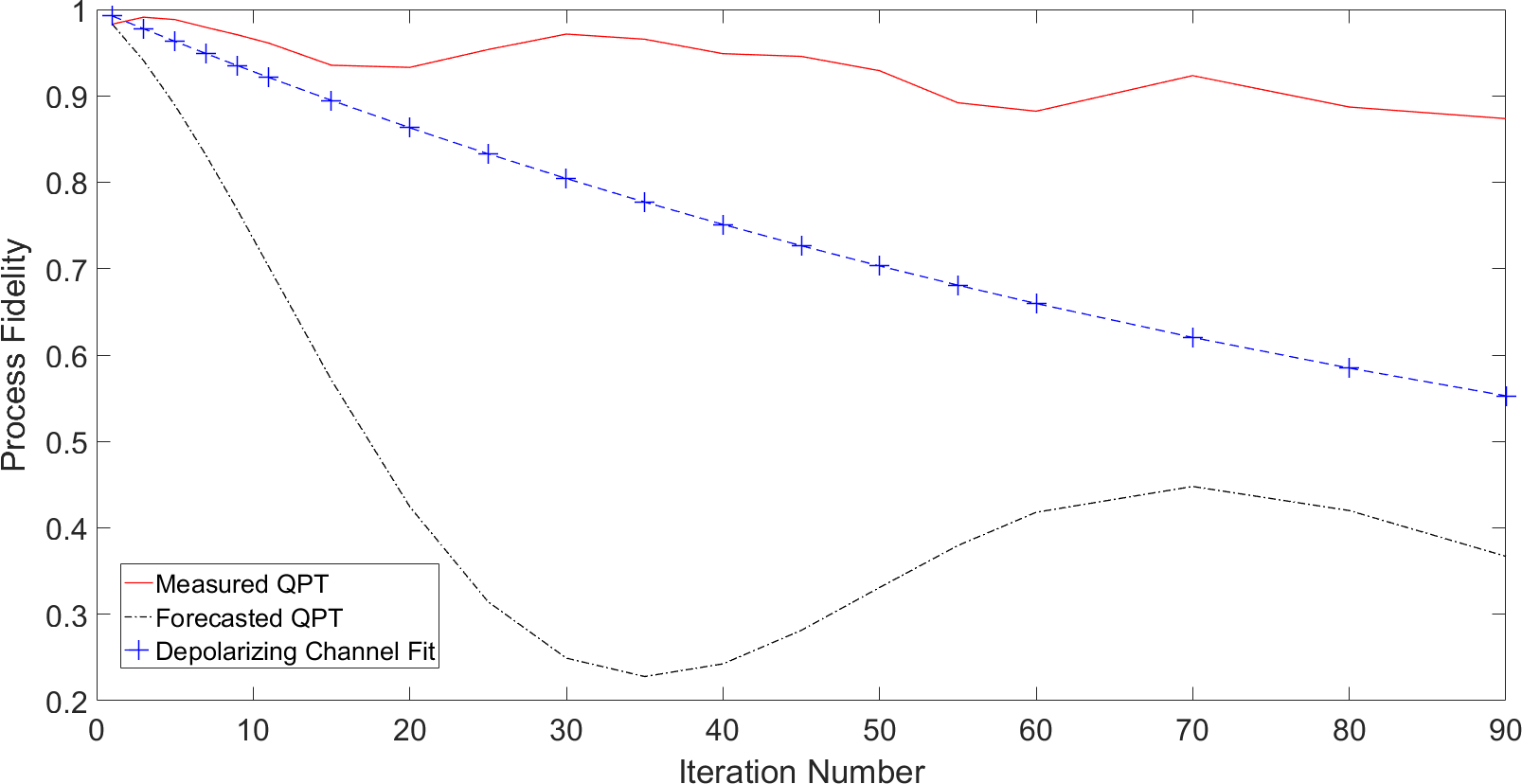}
\caption{Plot of experimentally measured process fidelity (determined via QPT) versus iteration count (red), along with forecasted results based on direct propagation of initially estimated chi matrix (black line) and depolarizing channel model (blue).}
\label{fig:forecasted}
\end{figure}

Using our device, we explored the behavior of the measured process fidelity upon performing multiple iterations of the identity gate, using QPT to estimate the fidelity for each iteration.  The results are summarized in Fig.\ \ref{fig:forecasted}. Surprisingly, the process fidelity after 90 iterations drops only to about 0.874.  This is well above the 0.55 cumulative process fidelity that is predicted from a simple fit to a depolarizing channel based on the fidelity of a single gate operation, which has a parameter value of $p = 0.010$.  Using instead the actual chi matrix estimate, and iterating the corresponding quantum operation, yields a sequence of process fidelity values with a curious, oscillatory behavior, as shown in Fig.\ \ref{fig:forecasted}.  Initially, it seems, the fidelity drops sharply.  After the $35^{\rm th}$ iteration, however, it begins to climb up again, only to crest and fall once more after about the $70^{\rm th}$ iteration.  This shows the folly of extrapolation based on a single QPT estimate.

\begin{figure}
\centering
\includegraphics[width=\textwidth]{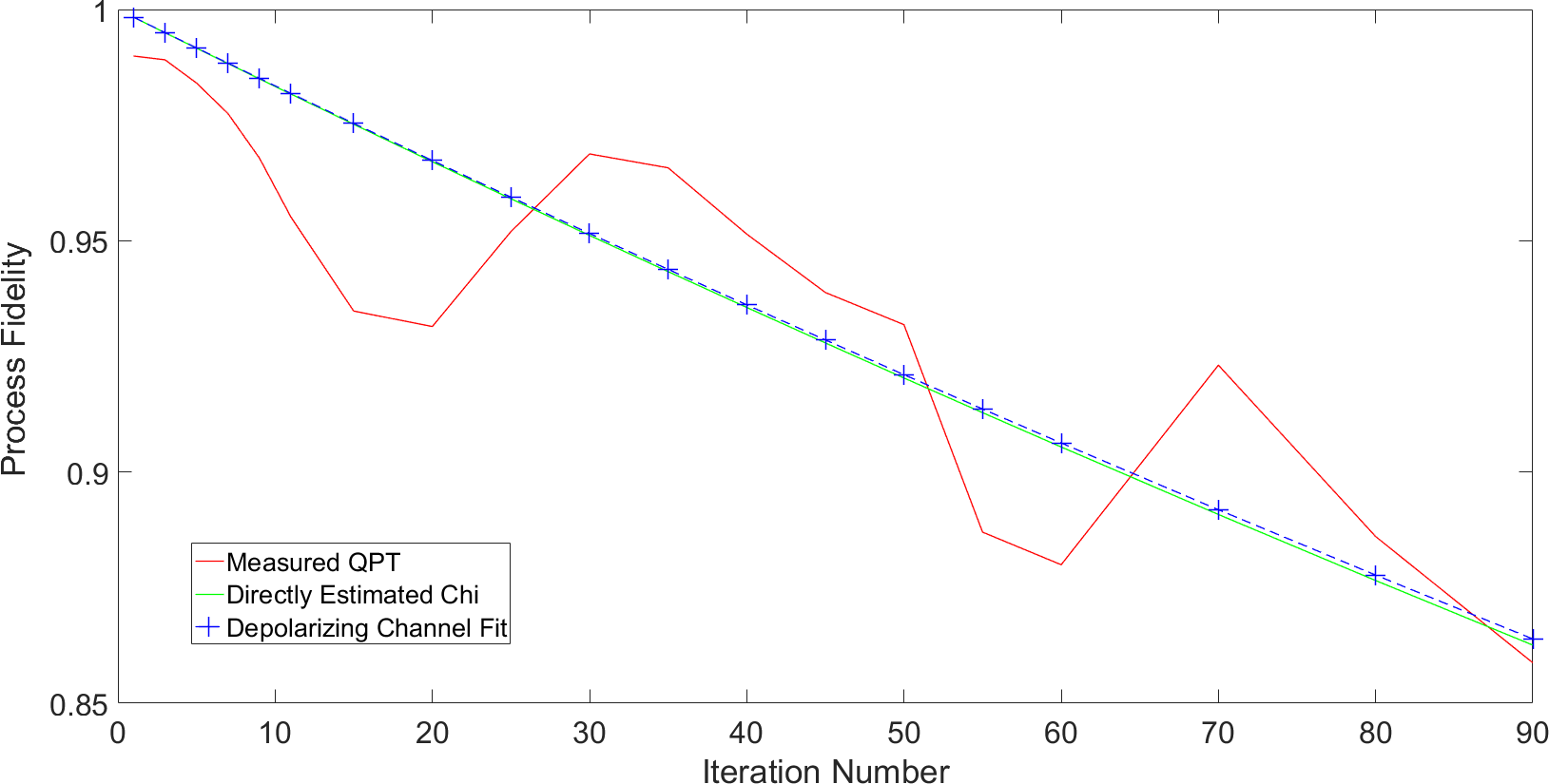}
\caption{Plot of experimentally measured process fidelity versus iteration count along with forecasted results based on fitting both an estimated chi matrix and a depolarizing channel model to the data using least squares fitting.}
\label{fig:fitted}
\end{figure}

If one instead considers the whole sequence of iterations, a much better fit can be achieved, as illustrated in Fig.\ \ref{fig:fitted}.  In this case, forecasting was done based on optimizing the channel parameterization in order to minimize the least-squares error to \emph{all} of the measured data (i.e., over all 90 iterations).  Doing so, we found that a simple depolarizing channel actually did provide a good fit to data, albeit with a model parameter value of $p = 0.006$.  This is just slightly lower than the value estimated from a single gate iteration, but the impact on the forecasted fidelity is quite significant.  We note that the resulting depolarizing channel also agrees quite well with the forecasted chi matrix, when fitted to all 90 iterations.

\section{Conclusions}
\label{sec:end}

We have considered the performance of a classical device in terms of quantum operations.  Using simple electronic hardware we are able to emulate the behavior of a two-qubit quantum computer by representing the quantum state as an analog voltage signal.  An arbitrary sequence of one- and two-qubit gate operations can be performed on these signals, thereby allowing for full programmability.  As with any physical device, these operations are not performed perfectly but, instead, exhibit some level of degradation.  In order to understand this better, we have chosen to use to the mathematical framework of quantum operations to model this classical degradation of performance in terms of quantum decoherence.  In particular, we avail ourselves of the techniques from the field of quantum state and quantum process tomography to extract an empirical estimate of the equivalent quantum channel corresponding to a given gate operation.

What we find is that estimate of the quantum channel obtained from performing QPT on a single gate operation provides a poor prediction of its performance upon repeated iterations.  If, however, the channel is estimated over a number of iterations, then a good fit can indeed be achieved.  This is particularly true when, as is the case for our device, the gate fidelity on any single iteration is quite high.  In such cases, the error is quite low and the estimation process will be very sensitive to the behavior over the first few iterations.  When these considerations are taken into account, we find that we are able to achieve a good fit to the measured results by assuming a simple depolarizing channel as a model for the effective quantum operation.

Given that decoherence may be viewed as a classical process that can be modeled quantum mechanically, the possibility arises for the use of quantum error correction techniques to improve performance.  In particular, we have shown that errors in a classical analog device can be modeled solely in terms of discrete bit-flip and phase error quantum operations, which are sufficient for modeling all forms of decoherence.  This presents the exciting new prospect of using the methods of fault-tolerant quantum computing to improve the fault tolerance of classical analog devices.  Whether the inclusion of additional, albeit faulty, classical resources can improve overall performance will be a subject for future investigations.


\begin{acknowledgement}
This work was support by the Office of Naval Research under Grant No. N00014-14-1-0323.
\end{acknowledgement}



\begin{thebibliography}{10}
\providecommand{\url}[1]{{#1}}
\providecommand{\urlprefix}{URL }
\expandafter\ifx\csname urlstyle\endcsname\relax
  \providecommand{\doi}[1]{DOI \discretionary{}{}{}#1}\else
  \providecommand{\doi}{DOI \discretionary{}{}{}\begingroup
  \urlstyle{rm}\Url}\fi

\bibitem{joos2013decoherence}
E.~Joos, H.D. Zeh, C.~Kiefer, D.J. Giulini, J.~Kupsch, I.O. Stamatescu,
  \emph{Decoherence and the appearance of a classical world in quantum theory}
  (Springer Science \& Business Media, 2013)

\bibitem{shor1995scheme}
P.W. Shor, Physical review A \textbf{52}(4), R2493 (1995)

\bibitem{gottesman2009introduction}
D.~Gottesman, in \emph{Quantum information science and its contributions to
  mathematics, Proceedings of Symposia in Applied Mathematics}, vol.~68 (2009),
  vol.~68, pp. 13--58

\bibitem{LaCour&Ott2015}
B.R. {La Cour}, G.E. Ott, New Journal of Physics \textbf{\textbf{17}}, 053017
  (2015)

\bibitem{LaCour&al2016}
B.R. {La Cour}, C.I. Ostrove, G.E. Ott, M.J. Starkey, G.R. Wilson,
  International Journal of Quantum Information \textbf{\textbf{14}}, 1640004
  (2016)

\bibitem{LaCour&Ostrove2017}
B.R. {La Cour}, C.I. Ostrove, Quantum Information Processing
  \textbf{\textbf{16}}, 7 (2017)

\bibitem{vonNeumann}
J.~{von Neumann}, \emph{Mathematical Foundations of Quantum Mechanics}
  (Princeton University Press, 1955)

\bibitem{Kwiat2006}
J.B. Altepeter, E.R. Jeffrey, P.G. Kwiat, \emph{Advances in Atomic, Molecular
  and Optical Physics} (Elsevier, 2006), vol. \textbf{52}, chap. Photonic State
  Tomography

\bibitem{Jozsa1994}
R.~Jozsa, Journal of Moderm Optics \textbf{\textbf{41}}, 2315 (1994)

\bibitem{Sudarshan1961}
E.C.G. Sudarshan, P.M. Mathews, J.~Rau, Physical Review \textbf{\textbf{121}},
  920 (1961)

\bibitem{Mike&Ike}
M.A. Nielsen, I.L. Chuang, \emph{Quantum Computation and Quantum Information}
  (Cambridge University Press, 2000)

\bibitem{Stinespring1955}
W.F. Stinespring, Proceedings of the American Mathematical Society
  \textbf{\textbf{6}}, 211 (1955)

\bibitem{Kraus}
K.~Kraus, \emph{States, Effects and Operations: Fundamental Notions of Quantum
  Theory} (Springer Verlag, 1983)

\bibitem{Choi1975}
M.D. Choi, Linear Algebra and its Applications \textbf{\textbf{10}}, 285 (1975)

\bibitem{Chuang1997}
I.L. Chuang, M.A. Nielsen, Journal of Modern Optics \textbf{\textbf{44}}, 2455
  (1997)

\bibitem{Poyatos1997}
J.~Poyatos, J.~Cirac, P.~Zoller, Physical Review Letters \textbf{\textbf{78}},
  390 (1997)

\bibitem{Bhandari2016}
R.~Bhandari, N.A. Peters, Scientific Reports \textbf{\textbf{6}}, 26004 (2016)

\bibitem{AGWhite2004mlqpt}
J.L. O'Brien, G.~Pryde, A.~Gilchrist, D.~James, N.K. Langford, T.~Ralph,
  A.~White, Physical review letters \textbf{93}(8), 080502 (2004)

\bibitem{Anis2012}
A.~Anis, A.I. Lvovsky, New Journal of Physics \textbf{\textbf{14}}, 105021
  (2012)

\bibitem{Yuen-Zhou2014}
J.~Yuen-Zhou, J.J. Krich, I.~Kassal, A.S. Johnson, A.~Aspuru-Guzik, in
  \emph{Ultrafast Spectroscopy}, 2053-2563 (IOP Publishing, 2014), pp. 1--1

\bibitem{gilchrist2005distance}
A.~Gilchrist, N.K. Langford, M.A. Nielsen, Physical Review A \textbf{71}(6),
  062310 (2005)

\end{thebibliography}

\end{document}